\documentclass[aps,prl,twocolumn,showpacs,preprintnumbers,amsmath,amssymb,superscriptaddress]{revtex4}%

\usepackage{graphicx}
\usepackage{dcolumn}
\usepackage{bm}
\usepackage{color}

\begin{document}

\preprint{preprint(\today)}

\title{Complementary response of static spin-stripe order and superconductivity to non-magnetic impurities in cuprates}

\author{Z.~Guguchia}
\email{zg2268@columbia.edu} 
\affiliation{Laboratory for Muon Spin Spectroscopy, Paul Scherrer Institute, CH-5232
Villigen PSI, Switzerland}
\affiliation{Department of Physics, Columbia University, New York, NY 10027, USA}

\author{B.~Roessli}
\affiliation{Laboratory for Neutron Scattering and Imaging, Paul Scherrer Institut, CH-5232 Villigen, Switzerland}

\author{R.~Khasanov}
\affiliation{Laboratory for Muon Spin Spectroscopy, Paul Scherrer Institute, CH-5232
Villigen PSI, Switzerland}

\author{A.~Amato}
\affiliation{Laboratory for Muon Spin Spectroscopy, Paul Scherrer Institute, CH-5232
Villigen PSI, Switzerland}

\author{E.~Pomjakushina}
\affiliation{Laboratory for Developments and Methods, Paul Scherrer Institut, CH-5232 Villigen PSI, Switzerland}

\author{K.~Conder}
\affiliation{Laboratory for Developments and Methods, Paul Scherrer Institut, CH-5232 Villigen PSI, Switzerland}

\author{Y.J.~Uemura}
\affiliation{Department of Physics, Columbia University, New York, NY 10027, USA}

\author{J.M.~Tranquada}
\affiliation{Condensed Matter Physics and Materials Science Division,
Brookhaven National Laboratory, Upton, NY 11973, USA}

\author{H.~Keller}
\affiliation{Physik-Institut der Universit\"{a}t Z\"{u}rich, Winterthurerstrasse 190, CH-8057 Z\"{u}rich, Switzerland}

\author{A.~Shengelaya}
\affiliation{Department of Physics, Tbilisi State University,
Chavchavadze 3, GE-0128 Tbilisi, Georgia}
\affiliation{Andronikashvili Institute of Physics of I.Javakhishvili Tbilisi State University,
Tamarashvili str. 6, 0177 Tbilisi, Georgia}

\begin{abstract}

We report muon-spin rotation and neutron-scattering experiments on non-magnetic Zn impurity effects on the static spin-stripe order and superconductivity of the La214 cuprates. Remarkably, it was found that, 
for samples with hole doping $x \approx 1/8$,
the spin-stripe ordering temperature $T_{so}$ decreases linearly with Zn doping $y$ and disappears at $y \approx 4\%$, demonstrating a high sensitivity of static spin-stripe order to impurities within a CuO$_{2}$ plane. Moreover, $T_{\rm so}$ is suppressed by Zn in the same manner as is the superconducting transition temperature $T_{\rm c}$ 
for samples near optimal hole doping.  This surprisingly similar sensitivity suggests that the spin-stripe order is dependent on intertwining with superconducting correlations.

\end{abstract}
\pacs{74.72.-h, 74.62.Fj, 75.30.Fv, 76.75.+i}
\maketitle


 One of the most astonishing manifestations of the competing ordered phases occurs in the system La$_{2-x}$Ba$_{x}$CuO$_{4}$ (LBCO) \cite{Bednorz}, where the bulk superconducting (SC) transition temperature $T_{\rm c}$ exhibits a deep minimum at $x$ = 1/8 \cite{Moodenbaugh,Kivelson,Vojta}. At this doping level muon-spin rotation (${\mu}$SR), neutron, and x-ray diffraction experiments revealed two-dimensional static charge and spin-stripe order  \cite{Tranquada1,Tranquada2,Abbamonte,Huecker,Luke,GuguchiaPRL,Fujita2004}.
The collected  experimental data indicate that the tendency toward uni-directional stripe-like ordering is common to cuprates \cite{Kivelson,Vojta,Kohsaka,Julien,Keimer2015}. However, the relevance of stripe correlations for high-temperature superconductivity remains a subject of controversy. On the theoretical front,  the concept of a sinusoidally-modulated pair-density wave (PDW) SC order, 
intimately intertwined with spatially modulated antiferromagnetism, has been introduced \cite{Berg1,Fradkin,Himeda}. On the experimental front, quasi-two-dimensional superconducting correlations were observed in La$_{1.875}$Ba$_{0.125}$CuO$_{4}$ (LBCO-1/8) and La$_{1.48}$Nd$_{0.4}$Sr$_{0.12}$CuO$_{4}$, coexisting with the ordering of static spin-stripes, but with frustrated phase order between the layers \cite{Tranquadareview,Tranquada2008,Li,Valla,Shen,Ding2008}.  Recently, it was found that in La$_{2-x}$Ba$_{x}$CuO$_{4}$ (0.11 ${\leq}$ $x$ ${\leq}$ 0.17) the 2D SC transition temperature $T_{\rm c1}$ and the static spin-stripe order temperature $T_{\rm so}$ have very similar values throughout the phase diagram \cite{Guguchiaarxiv,GuguchiaNJP}. Moreover, a similar pressure evolution of  $T_{\rm c}$ and $T_{\rm so}$ in the stripe phase of $x$ = 0.155 and 0.17 samples was observed.  These findings were discussed in terms of a spatially modulated and intertwined pair wave function \cite{Berg1,Fradkin,Himeda,Xu2014}. There are also a few reports proposing the relevance of a PDW state in sufficiently underdoped La$_{2-x}$Sr$_{x}$CuO$_{4}$ \cite{Jakobsen2015} and YBa$_{2}$Cu$_{3}$O$_{6-x}$ \cite{Lee,Yu}.  At present it is still unclear to what extent PDW order is a common feature of cuprate systems  where stripe order occurs.

\begin{figure}[b!]
\centering
\includegraphics[width=1.0\linewidth]{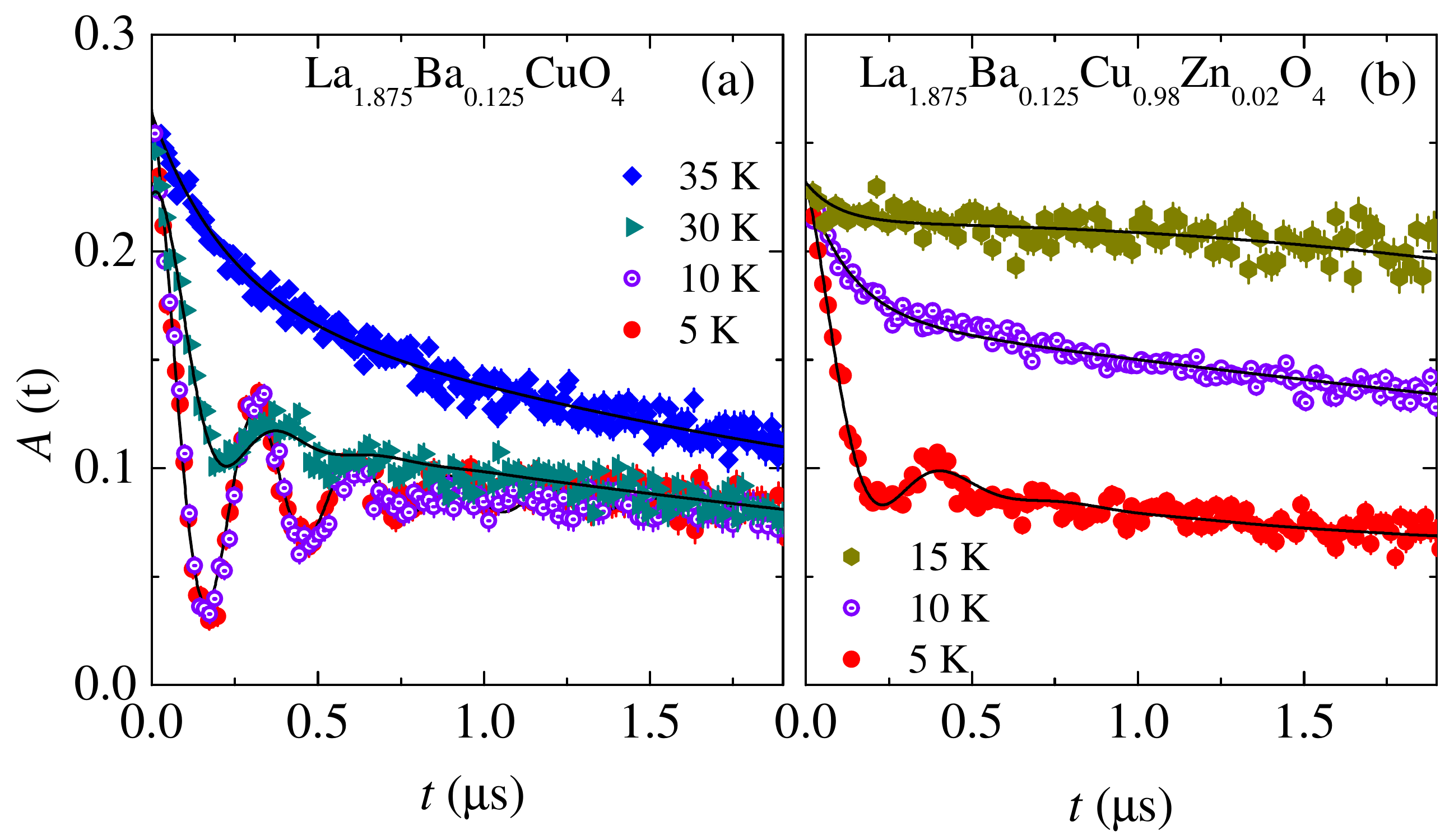}
\vspace{-0.6cm}
\caption{ (Color online) Zero-field (ZF) ${\mu}$SR time spectra $A(t)$ for La$_{1.875}$Ba$_{0.125}$CuO$_{4}$ (a) and La$_{1.875}$Ba$_{0.125}$Cu$_{0.98}$Zn$_{0.02}$O$_{4}$ (b) recorded at various temperatures. The solid lines represent fits to the data by means of Eq.~(3) of methods section.} 
\label{fig1}
\end{figure}

\begin{figure*}[t!]
\centering
\includegraphics[width=1.0\linewidth]{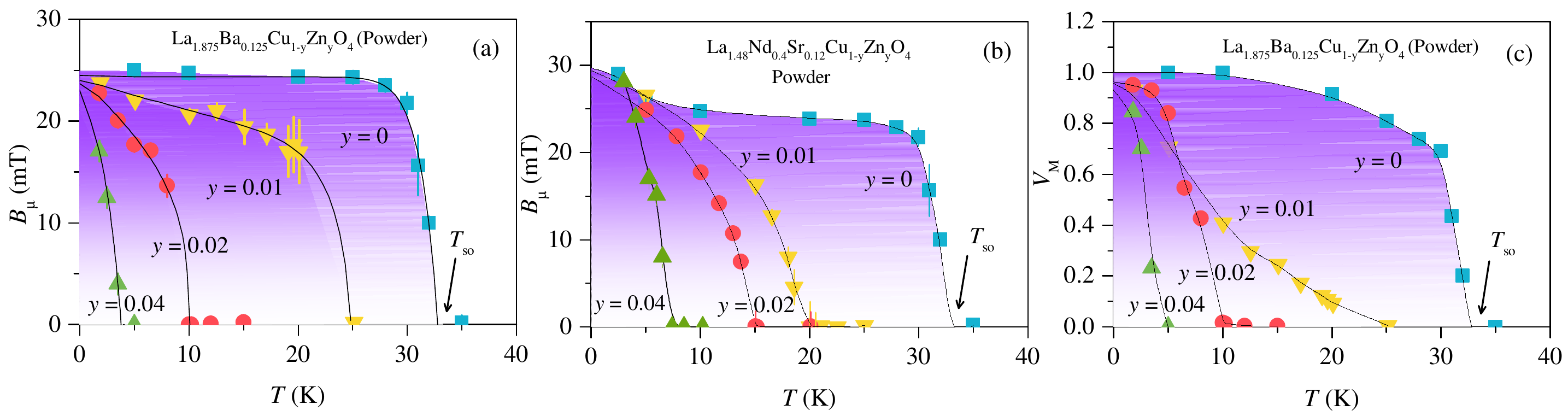}
\vspace{-0.7cm}
\caption{ (Color online) The temperature dependence of the internal magnetic field $B_{\rm \mu}$ (a) and the magnetic fraction $V_{\rm m}$  (c) for the polycrystalline samples of La$_{1.875}$Ba$_{0.125}$Cu$_{1-y}$Zn$_{y}$O$_{4}$ ($y$ = 0, 0.02, 0.04). The temperature dependence of  $B_{\rm \mu}$ for the polycrystalline samples of La$_{1.48}$Nd$_{0.4}$Sr$_{0.12}$Cu$_{1-y}$Zn$_{y}$O$_{4}$ ($y$ = 0, 0.01, 0.02, 0.04) (b). The arrows mark the spin-stripe order temperature $T_{\rm so}$. The solid curves are fits of the data to the power law $B_{\mu}$($T$) = $B_{\mu}$(0)[1-($T/T_{\rm so}$)$^{\gamma}$]$^{\delta}$, where $B_{\rm \mu}$(0) is the zero-temperature value of $B_{\mu}$. ${\gamma}$ and ${\delta}$ are phenomenological exponents.} 
\label{fig1}
\end{figure*}

 To further explore the interplay between static stripe order and superconductivity in cuprates we used non-magnetic impurity substitution at the Cu site as an alternative way of tuning the physical properties. Since the discovery of cuprate HTSs much effort was invested in the investigation of the effect of in-plane impurities. It is now well established that in cuprate HTSs nonmagnetic Zn ions suppress $T_{c}$ even more strongly than magnetic ions \cite{XiaoG,Fukuzumi,Komiya}. Such behaviour is in sharp contrast to that of conventional superconductors. This observation led to the formulation of an unconventional pairing mechanism and symmetry of the order parameter for cuprate HTSs. In addition, in several cases a ground state with static  antiferromagnetic (AF) correlations is stabilized by Zn-doping \cite{Mendels1994,Akoshima2000,Watanabe2000,Adachi2004,Adachi2004v2,Koike2012}. Up to now much less is known concerning impurity effects on the static stripe phase in cuprates at 1/8 doping.  From specific heat and neutron scattering measurements it was inferred that Zn doping leads to stripe destruction \cite{AnegawaZn,Takeda,Fujita}. Such an effect is very interesting and it was not predicted theoretically. However, no systematic impurity effect studies on static stripe order have been carried out up to now. Moreover, specific heat is a very indirect method to characterize the stripe phase in cuprates. Therefore, it is very important to use experimental techniques which can directly probe stripe formation and its evolution with impurity doping. 
   
  In this letter, we  report on systematic muon-spin rotation ${\mu}$SR, neutron scattering, and magnetization studies of Zn impurity effects on the static spin-stripe order and superconductivity in the La214 cuprates. Remarkably, it was found that in these systems the spin-stripe ordering temperature $T_{\rm so}$ decreases linearly with Zn doping $y$ and disappears at $y{\simeq}4\%$. This means that $T_{\rm so}$ is suppressed in the same manner as the superconducting transition temperature $T_c$ by Zn impurities. These results suggest that the stripe and SC orders may have a common physical mechanism and are intertwined.

 In a ${\mu}$SR experiment, positive muons implanted into a sample serve as an extremely sensitive local probe to detect small internal magnetic fields and ordered magnetic volume fractions in the bulk of magnetic systems. Thus ${\mu}$SR is a particularly powerful tool to study inhomogeneous magnetism in materials \cite{suppl}.  
 Neutron diffraction  experiments \cite{Eiger} allow to directly probe the incommensurate spin structure of spin-stripe order and thus provide crucial complementary information to the ${\mu}$SR technique.

\begin{figure}[b!]
\includegraphics[width=1.0\linewidth]{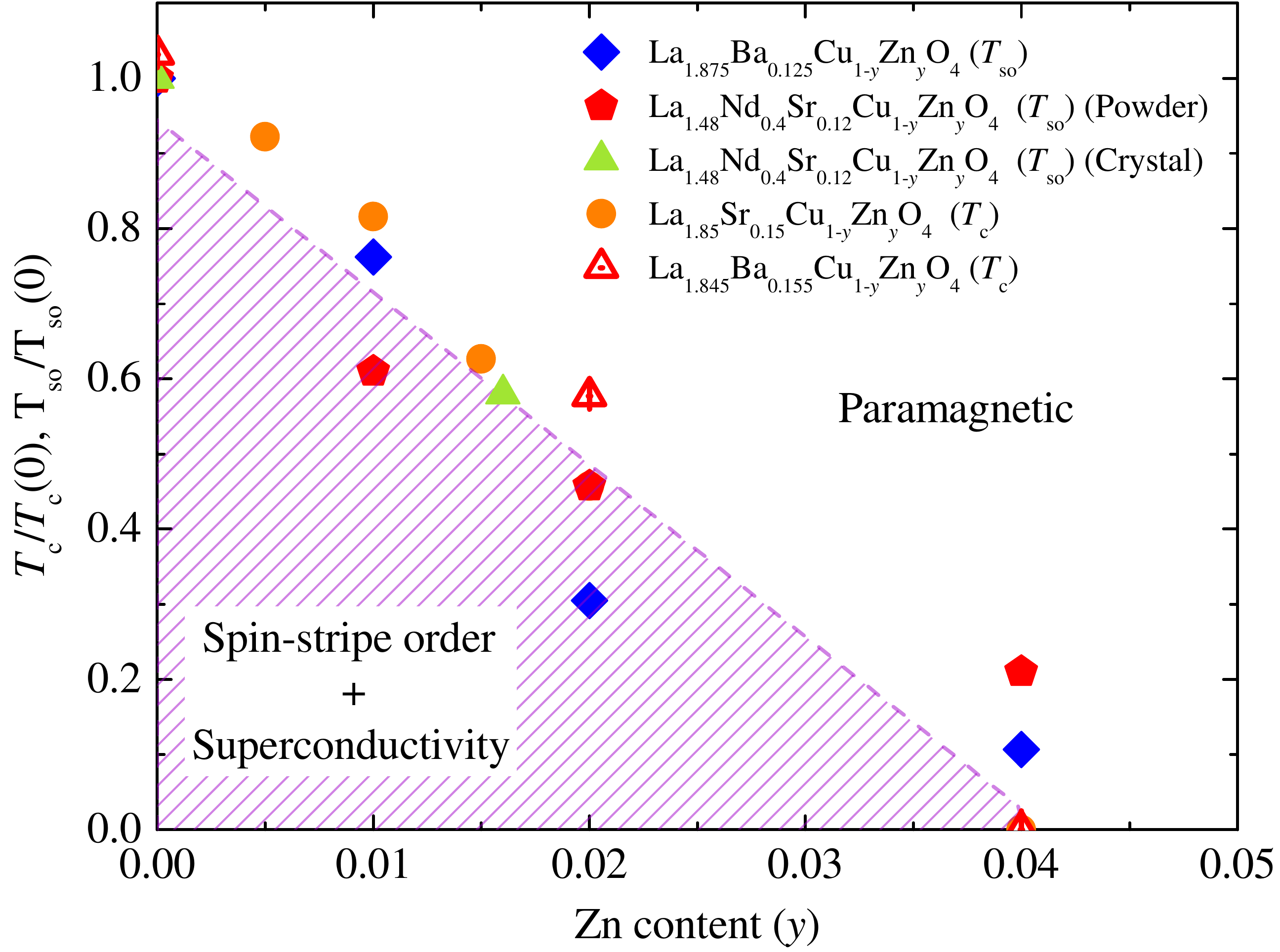}
\vspace{-0.5cm}
\caption{ (Color online) The normalised static spin-stripe order temperature $T_{\rm so}$/$T_{\rm so}$(0) for La$_{1.875}$Ba$_{0.125}$Cu$_{1-y}$Zn$_{y}$O$_{4}$ and La$_{1.48}$Nd$_{0.4}$Sr$_{0.12}$Cu$_{1-y}$Zn$_{y}$O$_{4}$ as a function of Zn content $y$. The superconducting transition temperature $T_{\rm c}$/$T_{\rm c}$(0) of La$_{1.85}$Sr$_{0.15}$Cu$_{1-y}$Zn$_{y}$O$_{4}$ as a function of Zn content $y$. The dashed line is a guide to the eye.}  
\label{fig1}
\end{figure}

\begin{figure*}[t!]
\centering
\includegraphics[width=0.8\linewidth]{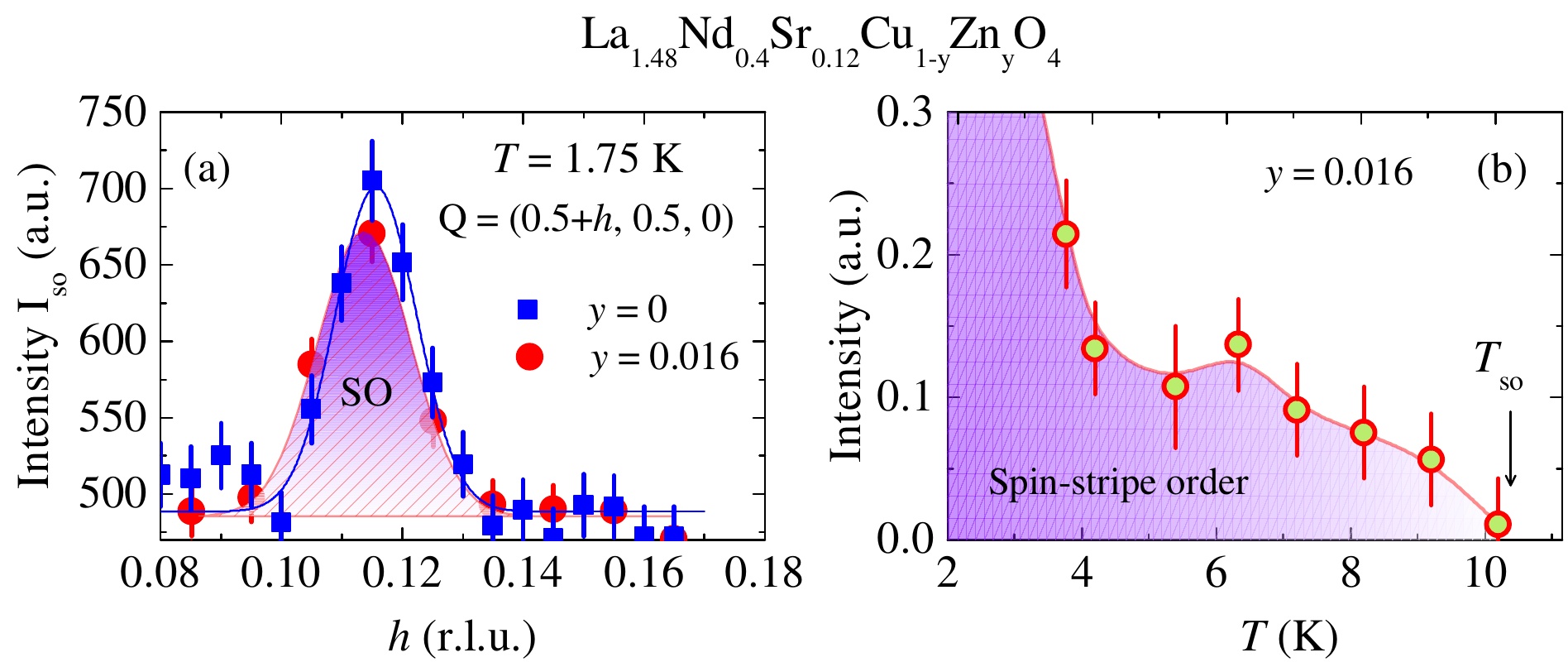}
\vspace{-0.5cm}
\caption{ (Color online) (a) $h$-scans through the SO-peak at (0.5+$h$,0.5,0) for the single crystals of La$_{1.48}$Nd$_{0.4}$Sr$_{0.12}$Cu$_{1-y}$Zn$_{y}$O$_{4}$ ($y$ = 0, 0.016), recorded at the base temperature $T$ = 1.75 K. The intensities have been normalized to the crystal
volume in the neutron beam. The solid lines represent the Gaussian fits to the data. (b) Peak intensity versus temperature of the (0.5+$h$,0.5,0) SO-peak, normalized to the crystal volume in the neutron beam.}
\label{fig1}
\end{figure*}

 Figures~1(a) and (b) show representative zero-field (ZF) ${\mu}$SR time spectra for polycrystalline La$_{1.875}$Ba$_{0.125}$Cu$_{1-y}$Zn$_{y}$O$_{4}$ samples with $y = 0$ and 0.02, respectively, recorded at various temperatures. For $y = 0$, damped oscillations due to muon-spin precession in internal magnetic fields are observed below $T_{\rm so}\approx35$~K, indicating the formation of static spin order in the stripe phase \cite{Luke,Arai,Nachumi,GuguchiaNJP,AndreasSuter,Crawford}. It is seen in Fig.~1(b), that for the $y = 0.02$ sample the oscillating signal appears only below $T\approx10$~K, showing strong suppression of the static spin-stripe order with Zn doping. We have studied this novel effect systematically as a function of Zn doping.

 The temperature dependence of the average internal field $B_{\rm \mu}$, which is proportional to the ordered magnetic moment, is shown in Fig.~2(a) for various Zn dopings $y$.  As evident from Fig.~2(a), $B_{\mu}(0)$, the internal magnetic field extrapolated to zero-temperature, does not depend on the Zn content $y$, while 
 $T_{\rm so}$ changes substantially with increasing $y$. 
Specifically, $T_{\rm so}$ decreases from $T_{\rm so}\simeq32.5$~K for $y = 0$ to $T_{\rm so}\simeq4$~K for $y = 0.04$. Figure 2(b) shows that a very similar behavior is observed for $B_{\rm \mu}$ measured on polycrystalline samples of the related compound La$_{1.48}$Nd$_{0.4}$Sr$_{0.12}$Cu$_{1-y}$Zn$_{y}$O$_{4}$. Note that the low-temperature value of $B_{\rm \mu}$ is enhanced by the ordering of the Nd moments. A similar suppression of $T_{\rm so}$ by Zn impurities was also observed in single-crystal samples of La$_{1.48}$Nd$_{0.4}$Sr$_{0.12}$Cu$_{1-y}$Zn$_{y}$O$_{4}$ ($y = 0$, 0.016). 
It seems that this effect is a generic feature of cuprates with static stripe order.
We note that in all the above mentioned systems, despite the suppression of $T_{\rm so}$ with Zn doping, the magnetic volume fraction $V_{\rm m}$ at the base temperature stays nearly 100\%\ [see Fig.~2(c)].
The bulk LTT structural phase transition temperature also stays nearly uneffected by Zn-doping (see supplementary Note II and supplementary Fig.~S1 \cite{suppl}).
  
  The observed Zn impurity effects on 
$T_{\rm so}$ in La$_{1.875}$Ba$_{0.125}$Cu$_{1-y}$Zn$_{y}$O$_{4}$ and La$_{1.48}$Nd$_{0.4}$Sr$_{0.12}$Cu$_{1-y}$Zn$_{y}$O$_{4}$ are summarised in Fig.~3. It is a remarkable finding that $T_{\rm so}$ linearly decreases with increasing Zn content $y$. Such a behaviour is reminiscent of the well known linear suppression of the SC transition temperature $T_{\rm c}$ in cuprates \cite{XiaoG,Fukuzumi,Komiya}. Since the superconducting volume fraction in 1/8 doped samples is tiny and the bulk $T_{\rm c}$ is also very low, it is difficult to follow the SC properties of these systems as a function of Zn content.
Alternatively, in Fig.~3 we plot $T_{\rm c}$ values for  optimally doped La$_{1.85}$Sr$_{0.15}$CuO$_{4}$ \cite{XiaoG,Fukuzumi,Komiya} and La$_{1.845}$Ba$_{0.155}$CuO$_{4}$ (see below) as a function of Zn content. Strikingly, 
suppression of $T_{\rm so}$ goes in a very similar manner as the well known impurity-induced $T_c$ suppression.  
  
 We have confirmed the Zn doping effect on the static spin-stripe order by neutron diffraction experiments
on single-crystal samples of La$_{1.48}$Nd$_{0.4}$Sr$_{0.12}$Cu$_{1-y}$Zn$_{y}$O$_{4}$ ($y = 0$, 0.016) \cite{composition}. 
The magnetic ordering wave vectors are ${\bf Q}_{so} = (0.5-\delta,0.5,0)$ and $(0.5,0.5+\delta,0)$, i.e., they are displaced by ${\delta}$ from the position of the magnetic Bragg peak in the AF parent compound La$_{2}$CuO$_{4}$ \cite{Tranquada1,Tranquada2}. In Fig. 4(a) we show $h$ scans through the $(0.5 + \delta,0.5,0)$ magnetic superlattice peaks, recorded at $T = 1.75$~K for the samples $y = 0$ and 0.016. It is clear that the intensity and incommensurability do not change with Zn doping. However, $T_{\rm so}$ is strongly suppressed from $T_{\rm so}\simeq50$~K \cite{Tranquada1}  for $y = 0$ to $T_{\rm so}\simeq10$~K for $y = 0.016$, as demonstrated in Fig.~4(b), where the peak intensity is shown as a function of temperature.
  
Going further, we studied the Zn-impurity effects on $T_{\rm so}$ and $T_{\rm c}$ in La$_{1.845}$Ba$_{0.155}$CuO$_{4}$. This compound ($x$  ${\textgreater}$ 1/8) exhibits a well defined bulk SC transition with $T_{\rm c}$ = 30 K and at the same time shows static spin-stripe order $T_{\rm so}$ ${\simeq}$ $T_{\rm c}$ = 30 K \cite{Guguchiaarxiv}. This enables us to study impurity effects on $T_{\rm so}$ and $T_{\rm c}$ simultaneously in the same sample. Figure~5a shows the temperature dependence of the magnetic volume fraction $V_{m}$ extracted from ZF-${\mu}$SR data for La$_{1.845}$Ba$_{0.155}$Cu$_{1-y}$Zn$_{y}$O$_{4}$ ($y$ = 0, 0.02, and 0.04). The low temperature value of $V_{m}$ increases with increasing Zn content $y$ and reaches 100 ${\%}$ for the highest Zn content $y$ = 0.04. On the other hand, $T_{{\rm so}}$ decreases with increasing $y$ similar as for 1/8-doping. The values of $T_{{\rm so}}$ and $T_{{\rm c}}$ (see the supplementary Note III and supplementary Figs. S2 and S3 \cite{suppl}) as a function of Zn content $y$ are shown in Fig. 5(b). Again, with increasing $y$ both $T_{{\rm c}}$ and $T_{{\rm so}}$ decrease  linearly with the same slope, indicating that Zn impurities influence $T_{{\rm c}}$ and $T_{{\rm so}}$ in the same manner. 
 
\begin{figure}[t!]
\includegraphics[width=1.0\linewidth]{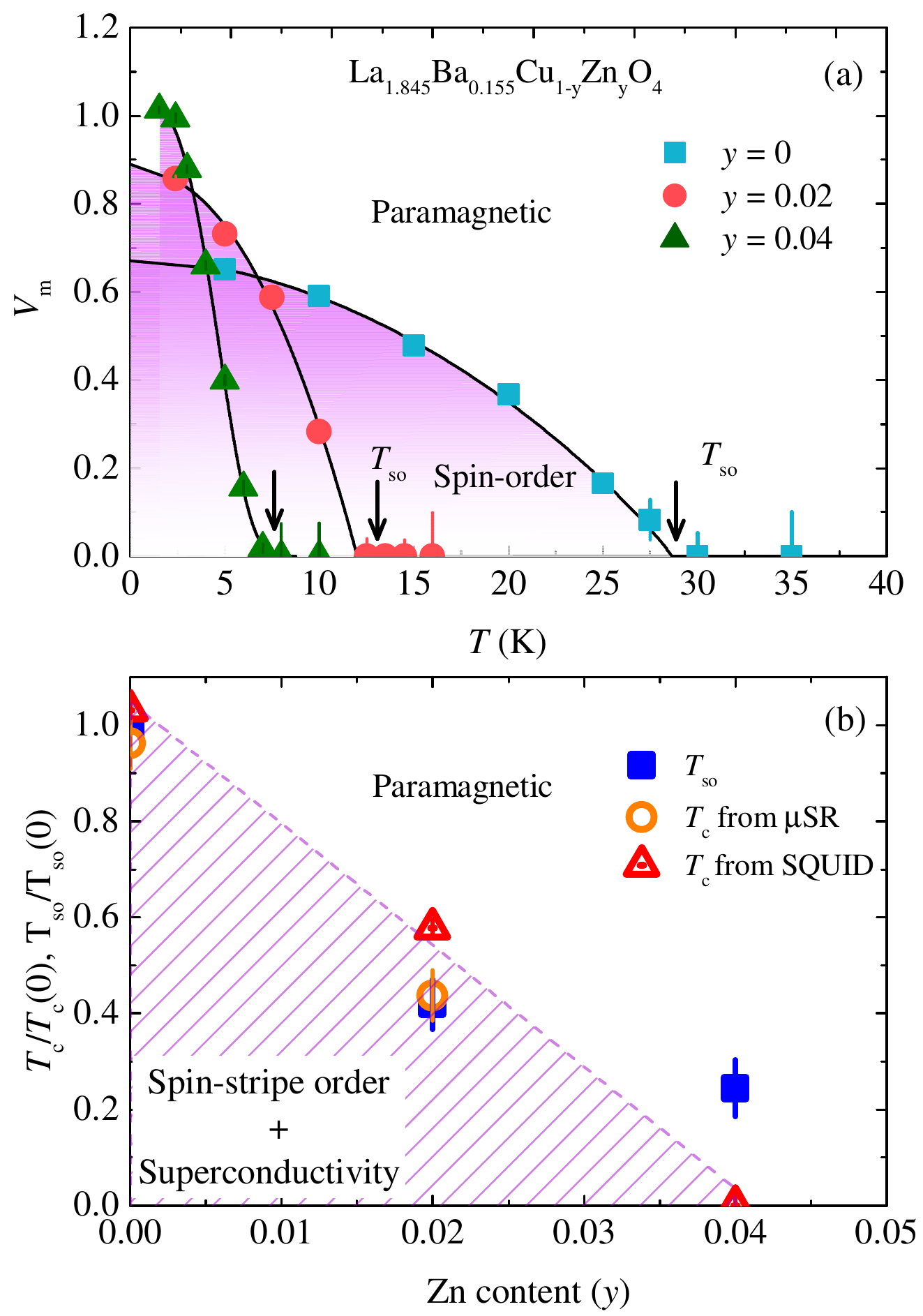}
\vspace{-0.5cm}
\caption{ (Color online) (a) Temperature dependence of the magnetic volume fraction $V_{\rm m}$ for La$_{1.845}$Ba$_{0.155}$Cu$_{1-y}$Zn$_{y}$O$_{4}$ (y = 0, 0.02, 0.04). (b) The normalized static spin-stripe order temperature $T_{\rm so}$/$T_{\rm so}$(0) and  the normalized superconducting transition temperature $T_{\rm c}$/$T_{\rm c}$(0) for La$_{1.845}$Ba$_{0.155}$Cu$_{1-y}$Zn$_{y}$O$_{4}$ as a function of Zn content $y$. The dashed line is a guide to the eye.}  
\label{fig1}
\end{figure}
  
 What is the significance of this surprising correlation?  Let us start with the fact that it is unusual to have spin order occur in a hole-doped cuprate at a temperature of $\sim 35$~K.  Just a couple of percent of hole doping is generally sufficient to wipe out antiferromagnetic order \cite{Matsuda2002}.  One common point of view is that antiferromagnetism and superconductivity are competing orders \cite{Sachdev2000}.  From that perspective, one might take the occurrence of spin-stripe order as evidence that hole-pairing and superconductivity have been suppressed.  In that case, we might expect the impact of Zn-doping on $T_{\rm so}$ to be similar to its impact on the N\'eel temperature in La$_2$CuO$_4$. That assumption leads to a problem, however, as experiment has demonstrated that it takes, not 4\%, but $\sim40\%$ Zn to destroy N\'eel order \cite{Vajk2002}.   One could also take account of the fact that the Zn tends to induce static Cu spin order in its immediate neighborhood \cite{Mahajan1994,Julien2000}, which, with random locations of the Zn sites, could lead, at higher Zn concentrations, to some disorder from neighboring pinned stripe domains being out of phase with one another; however, a shortening of the spin correlation length only becomes apparent with at least 3\%\ Zn doping \cite{Fujita,Birgenau}, while the drop in $T_{\rm so}$ is clear at much lower Zn concentrations.

 Consider instead that previous experiments provide evidence that spin-stripe order coexists with two-dimensional superconducting correlations in LBCO-1/8 \cite{Tranquada2008,Li}.  Here, the superconducting and spin orders are intertwined \cite{Fradkin}.   Superconducting correlations within charge stripes must establish Josephson coupling across the spin stripes, while the spins in neighboring stripes must establish an effective exchange coupling via the fluctuating pairs in the intervening charge stripe.  A Zn ion will locally suppress hole motion, thus eliminating local superconducting coherence and weakening the superconductivity \cite{Balatsky}.  Local suppression of hole hopping will also disrupt the effective exchange coupling between spin stripes, leading to a reduction in $T_{\rm so}$.

Previous $\mu$SR studies of Zn doping in LSCO and YBa$_{2}$Cu$_{3}$O$_{7}$ have established the ``Swiss-cheese`` model: a fixed carrier density per Zn atom is removed from the superfluid density, as if each Zn removes a fixed areal fraction of the superfluid \cite{Nachumi1996}.  The linear relationship between $T_c$ and the average superfluid density, valid for underdoped through optimally-doped cuprate HTSs, then explains the reduction of $T_c$ with increasing Zn concentration \cite{Uemura1989}.  For the stripe-ordered systems, it is plausible that both the superconducting and spin-stripe orders will respond in a similar fashion.

  In conclusion, static spin-stripe order and superconductivity in cuprate systems La$_{2-x}$Ba$_{x}$Cu$_{1-y}$Zn$_{y}$O$_{4}$ ($x$ = 0.125, 0.155) and La$_{1.48}$Nd$_{0.4}$Sr$_{0.12}$Cu$_{1-y}$Zn$_{y}$O$_{4}$ were studied by means of magnetisation, ${\mu}$SR, and neutron scattering experiments as a function of nonmagnetic Zn impurity concentration. High sensitivity of the static spin-stripe order temperature $T_{\rm so}$ to impurities in the CuO$_{2}$ plane was demonstrated. Namely, the spin-stripe ordering temperature $T_{\rm so}$ strongly decreases linearly with Zn doping and disappears at about 4\%\ Zn content. More strikingly, $T_{\rm so}$ is suppressed in the same fashion as is the superconducting transition temperature $T_{\rm c}$ by Zn impurities.  These results strongly suggest that the existence of the stripe order requires intertwining with the SC pairing correlations, such as occurs in the proposed PDW state. The present findings should help to better understand the complex interplay between stripe order and superconductivity in cuprates. More generally, since charge and spin orders are often observed in other transition-metal oxides, investigation of impurity effects and disorder on stripe formation may become an interesting research avenue in correlated electron systems.

\textit{Acknowledgments.}  The ${\mu}$SR experiments were carried out at the ${\pi}$M3 beam line of the Paul Scherrer Institute (Switzerland), using the general purpose instrument (GPS). The neutron scattering experiments were carried out with the three-axis spectrometer EIGER at the Swiss Spallation Neutron Source SINQ at the Paul Scherrer Institut (PSI), Switzerland. We are grateful to S.A. Kivelson for valuable discussions. Z.G. gratefully acknowledges the financial support by the Swiss National Science Foundation (Early Postdoc Mobility SNFfellowship P2ZHP2-161980 and SNFGrant 200021-149486). Z.G. thanks Martin Mansson for useful discussions. 
A.S. acknowledges support from the SCOPES grant No. SCOPES IZ74Z0-160484.
Work at Columbia University is supported by US NSF DMR-1436095 (DMREF) and NSF DMR-1610633 as well as by REIMEI project of Japan Atomic Energy Agency.
JMT is supported at Brookhaven by the U.S. Department of Energy, Office of Basic Energy Sciences, under contract No.\ DE-SC0012704.



\newpage
 
\section{SUPPLEMENTAL MATERIAL}

\section{METHODS}

\textbf{Sample preparation}: Polycrystalline samples of La$_{1.875}$Ba$_{0.125}$Cu$_{1-y}$Zn$_{y}$O$_{4}$ with $y$ = 0, 0.02, 0.04 and La$_{1.48}$Nd$_{0.4}$Sr$_{0.12}$Cu$_{1-y}$Zn$_{y}$O$_{4}$ with $y$ = 0, 0.01, 0.02, 0.04 were prepared by the conventional solid-state reaction method using La$_{2}$O$_{3}$, Nd$_{2}$O$_{3}$, BaCO$_{3}$, SrCO$_{3}$, and CuO. The single-phase character of the samples was checked by powder x-ray diffraction. The single crystals of La$_{1.48}$Nd$_{0.4}$Sr$_{0.12}$Cu$_{1-y}$Zn$_{y}$O$_{4}$ ($y$ = 0, 0.016) were grown by the traveling solvent floating zone method. All the measurements were performed on samples from the same batch.\\
 
\begin{figure}[b!]
\includegraphics[width=0.8\linewidth]{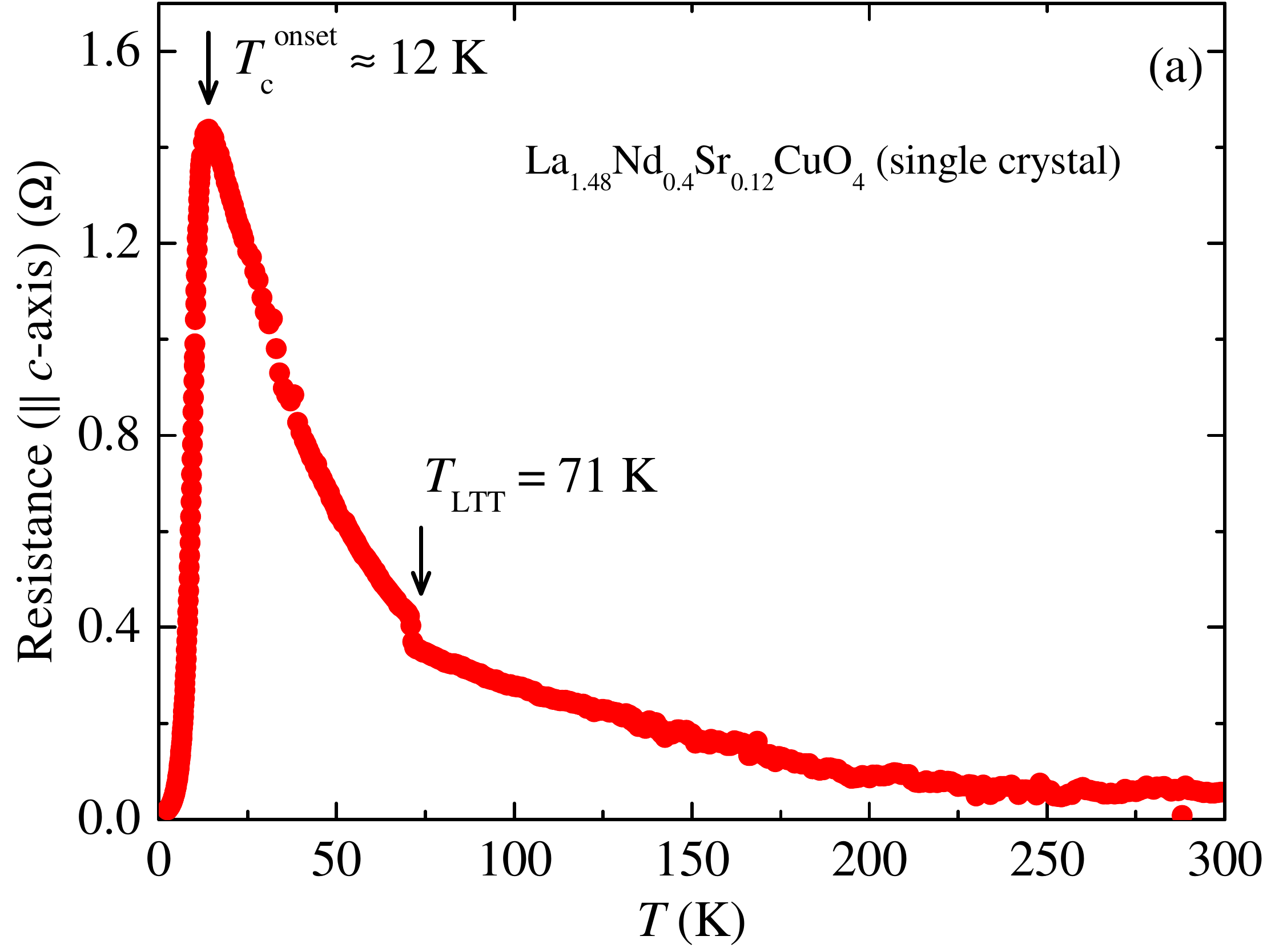}
\includegraphics[width=0.8\linewidth]{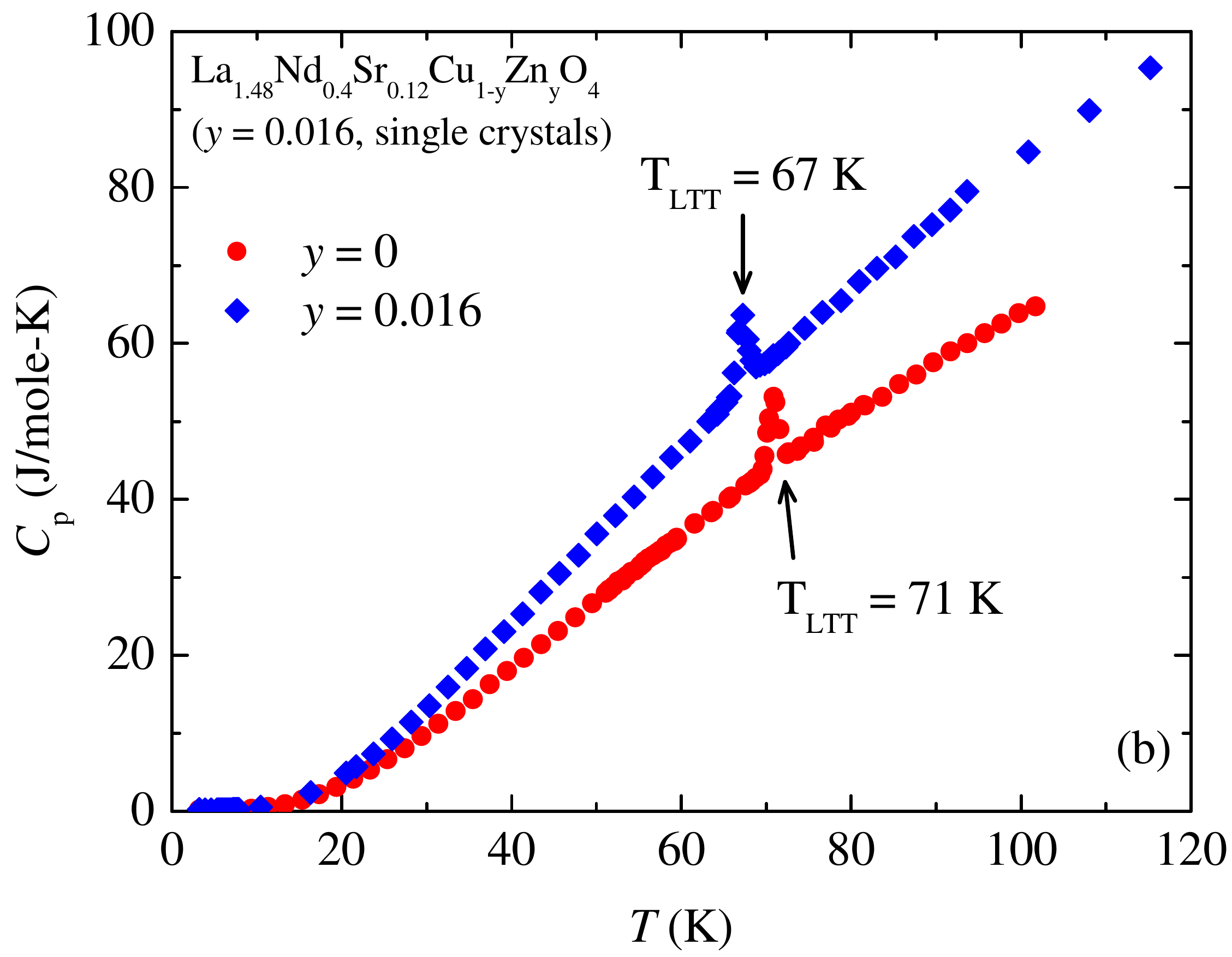}
\caption{ (Color online) (a) Temperature dependence of the resistance along the $c$-axis for the single crystal of La$_{1.48}$Nd$_{0.4}$Sr$_{0.12}$CuO$_{4}$.
(b) The specific heat $C_{\rm p}$ as a function of temperature for the single crystals of La$_{1.48}$Nd$_{0.4}$Sr$_{0.12}$Cu$_{1-y}$Zn$_{y}$O$_{4}$ ($y$ = 0, 0.016). The arrows denote the LTT structural phase transition temperatures $T_{\rm LTT}$.}
\label{fig1}
\end{figure}
 
\textbf{Principles of the ${\mu}$SR technique}: Static spin-stripe orders in La$_{1.875}$Ba$_{0.125}$Cu$_{1-y}$Zn$_{y}$O$_{4}$ with $y$ = 0, 0.02, 0.04 and La$_{1.48}$Nd$_{0.4}$Sr$_{0.12}$Cu$_{1-y}$Zn$_{y}$O$_{4}$ with $y$ = 0, 0.01, 0.02, 0.04 were studied by means of zero-field (ZF) ${\mu}$SR experiments.  The ${\mu}$SR experiments were carried out at the ${\pi}$M3 beam line of the Paul Scherrer Institute (Switzerland), using the general purpose instrument (GPS). In a ${\mu}$SR experiments an intense beam ($p_{\mu}$ = 29 MeV/c) of 100 ${\%}$ spin-polarized muons is stopped in the sample mounted inside of a gas-flow $^{4}$He cryostat on a sample holder with a standard veto setup providing
essentially a background free ${\mu}$SR signal. The positively charged muons thermalize in the sample at interstitial
lattice sites, where they act as magnetic microprobes. In a magnetic
material the muons spin precess in the local field $B_{{\rm \mu}}$
at the muon site with the Larmor frequency ${\nu}_{{\rm \mu}}$ =
$\gamma_{{\rm \mu}}$/(2${\pi})$$B_{{\rm \mu}}$ (muon gyromagnetic
ratio $\gamma_{{\rm \mu}}$/(2${\pi}$) = 135.5 MHz T$^{-1}$). 
In a ZF ${\mu}$SR experiment positive muons implanted into a sample serve as an extremely
sensitive local probe to detect small internal magnetic fields and ordered magnetic volume fractions
in the bulk of magnetic materials. Thus, ${\mu}$SR is a particularly powerful
tool to study inhomogeneous magnetism in materials.

\begin{figure*}[t!]
\centering
\includegraphics[width=1.0\linewidth]{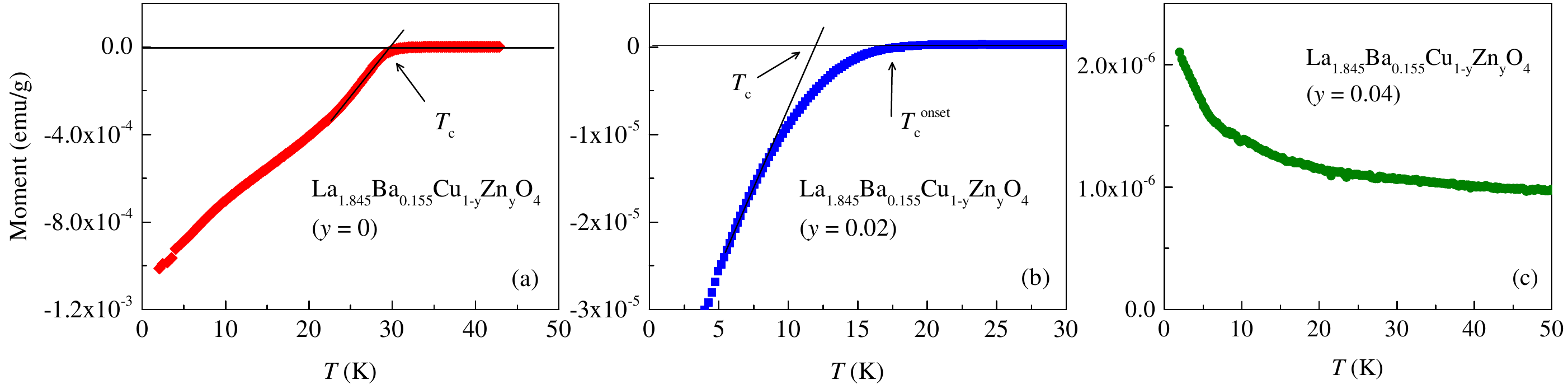}
\vspace{-0.7cm}
\caption{ (Color online) Temperature dependence of the diamagnetic moment of  La$_{1.845}$Ba$_{0.155}$Cu$_{1-y}$Zn$_{y}$O$_{4}$ for various $y$ = 0, 0.02 and 0.04, measured in a magnetic field of $\mu_{0}H$ = 0.5 mT. The arrows denote the superconducting transition temperature $T_{\rm c}$.} 
\label{fig1}
\end{figure*}
\begin{figure}[b!]
\includegraphics[width=0.7\linewidth]{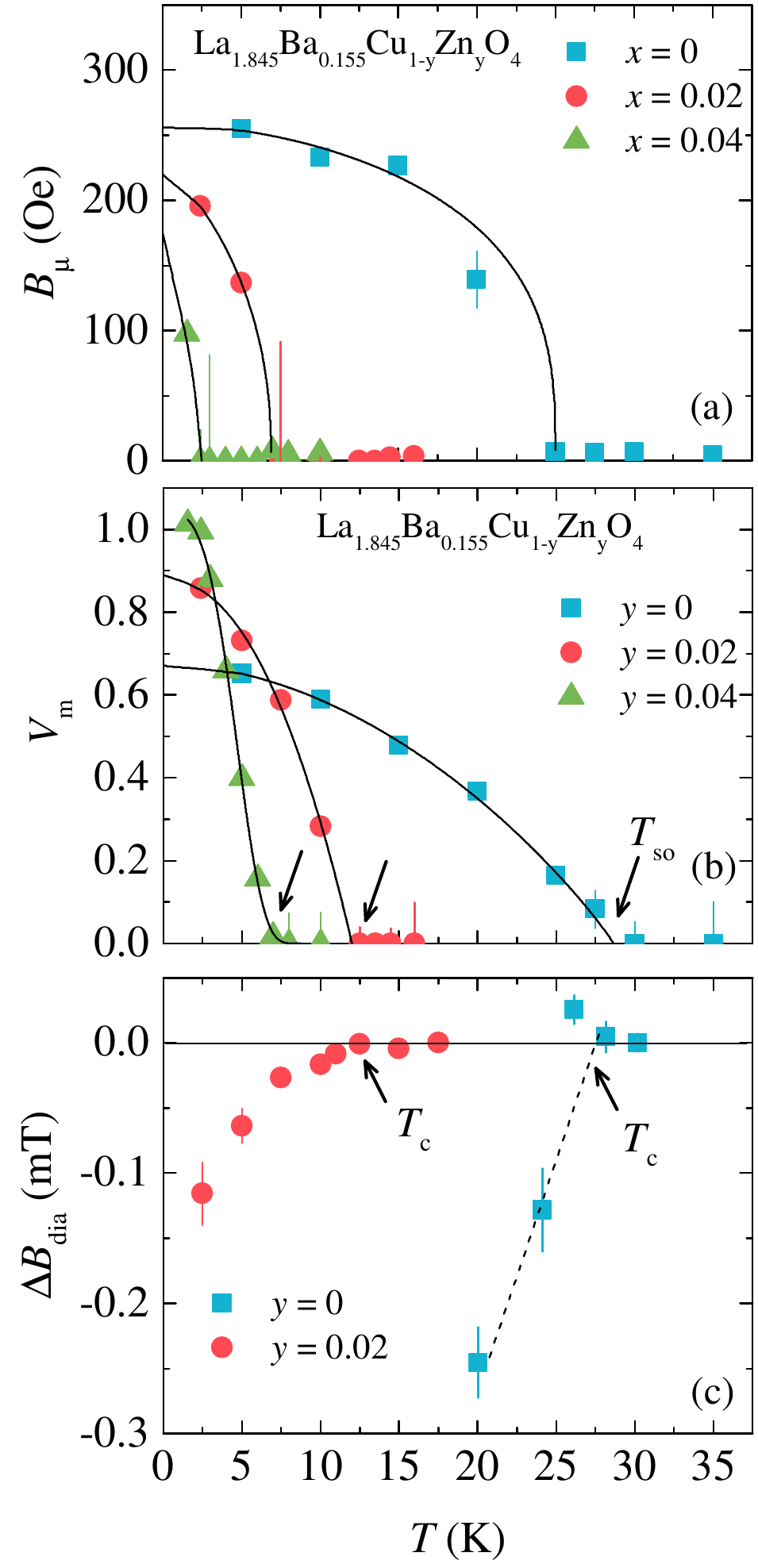}
\vspace{-0.3cm}
\caption{ (Color online) Temperature dependence of the internal magnetic field $B_{\mu}$ (a) and magnetic volume fraction $V_{m}$ (b) in  La$_{1.845}$Ba$_{0.155}$Cu$_{1-y}$Zn$_{y}$O$_{4}$ ($y$ = 0, 0.02, 0.04). The solid lines are fits of the data to the same empirical power law as described in the manuscript. (c) Diamagnetic shift ${\Delta}$$B_{{\rm dia}}$ of  La$_{1.845}$Ba$_{0.155}$Cu$_{1-y}$Zn$_{y}$O$_{4}$ ($y$ = 0, 0.02, 0.04) as a function of temperature. The arrows denote $T_{c}$.}
\end{figure}

The muons $\mu^{+}$ implanted into the sample will decay after a mean life time of ${\tau}$$_{\mu}$ = 2.2 ${\mu}$s, emitting a fast positron $e^{+}$  preferentially along their spin direction. Various detectors placed around the sample track the incoming $\mu^{+}$ and the outgoing $e^{+}$. When the muon detector records the arrival of a ${\mu}$ in the specimen,
the electronic clock starts. The clock is stopped when the decay positron $e^{+}$ is registered in
one of the $e^{+}$ detectors, and the measured time interval is stored in a histogramming memory.
In this way a positron-count versus time histogram is formed.
A muon decay event requires that within a certain time interval
after a $\mu^{+}$ has stopped in the sample a $e^{+}$ is detected. This time interval extends usually over several muon lifetimes ($e.g.$ 10${\mu}$s). After a bunch of muons stopped in the sample, one obtains a histogram 
for the forward ($N_{\rm e^{+}F}$) and the backward ($N_{\rm e^{+}B}$) detectors, 
which in the ideal case has the following form:

\begin{equation} 
N_{e^{+}{\alpha}(t)}=N_{0}e^{-\frac{t}{\tau_{\mu}}}(1+A_{0}\vec{P}(t)\hat{n}_{\alpha})+N_{bgr}. ~ {\alpha} = F, B
\end{equation}
 
 Here, the exponential factor accounts for the radioactive muon decay.
$\vec{P}$(t) is the muon-spin polarization function with the unit vector 
${\hat{n}_{\alpha}}$ (${\alpha}$ = F,B) with respect to the incoming muon spin polarization.
$N_{\rm 0}$ is number of positrons at the initial time $t$=0.    
$N_{\rm bgr}$ is a background contribution due to uncorrelated starts and stops.
$A_{0}$ is the initial asymmetry, depending on different experimental factors, such as the detector solid angle,
efficiency, absorption, and scattering of positrons in the material. Typical values of $A_{0}$ are
between 0.2 and 0.3.

  Since the positrons are emitted predominantly in the direction of the muon spin which precesses
with ${\omega_{\mu}}$, the forward and backward detectors will detect a signal oscillating
with the same frequency. In order to remove the exponential decay due to the finite 
life time of the muon, the so-called asymmetry signal $A$(t) is calculated:
\begin{equation} 
A(t)=\frac{N_{e^{+}F}(t)-N_{e^{+}B}(t)}{N_{e^{+}F}(t)+N_{e^{+}B}(t)}=A_{0}P(t),
\end{equation}
where, $N_{e^{+}F}$(t) and $N_{e^{+}B}$(t) are the number of positrons detected in the forward and backward 
detectors, respectively. The quantities $A(t)$ and $P(t)$ depend sensitively on the
spatial distribution and dynamical fluctuations of the magnetic environment of the muons.
Hence, these functions allow to study  interesting physics of the investigated system.\\

\textbf{Analysis of ZF-${\mu}$SR data}: 

The ${\mu}$SR signals in the whole temperature range were analyzed by decomposing
the signal into a magnetic and a nonmagnetic contribution:\\ 
\begin{equation}
P(t)=V_{m}\Bigg[{\frac{2}{3}e^{-\lambda_{T}t}J_0(\gamma_{\mu}B_{\mu}t)}+\frac{1}{3}e^{-\lambda_{L}t}\Bigg]
+(1-V_{m})e^{-\lambda_{nm}t}.
\label{eq1}
\end{equation}
Here, $V_{\rm m}$ denotes the relative volume of the magnetic fraction, and $B_{\mu}$ is the average internal magnetic field at the muon site. ${\lambda_T}$ and ${\lambda_L}$
are the depolarization rates representing the transversal and the longitudinal 
relaxing components of the magnetic parts of the sample.
$J_{0}$ is the zeroth-order Bessel function of the first kind.
This is characteristic for an incommensurate spin density wave 
and has been observed in cuprates with static spin stripe order \cite{Nachumi}.
${\lambda_{nm}}$ is the relaxation rate of the nonmagnetic part of the sample, where spin-stripe order is absent.
The ${\mu}$SR time spectra were analyzed using the free software package MUSRFIT \cite{AndreasSuter}.\\

\section{Studies of transport, thermodynamic and structural properties}

\begin{figure}[t!]
\includegraphics[width=1.0\linewidth]{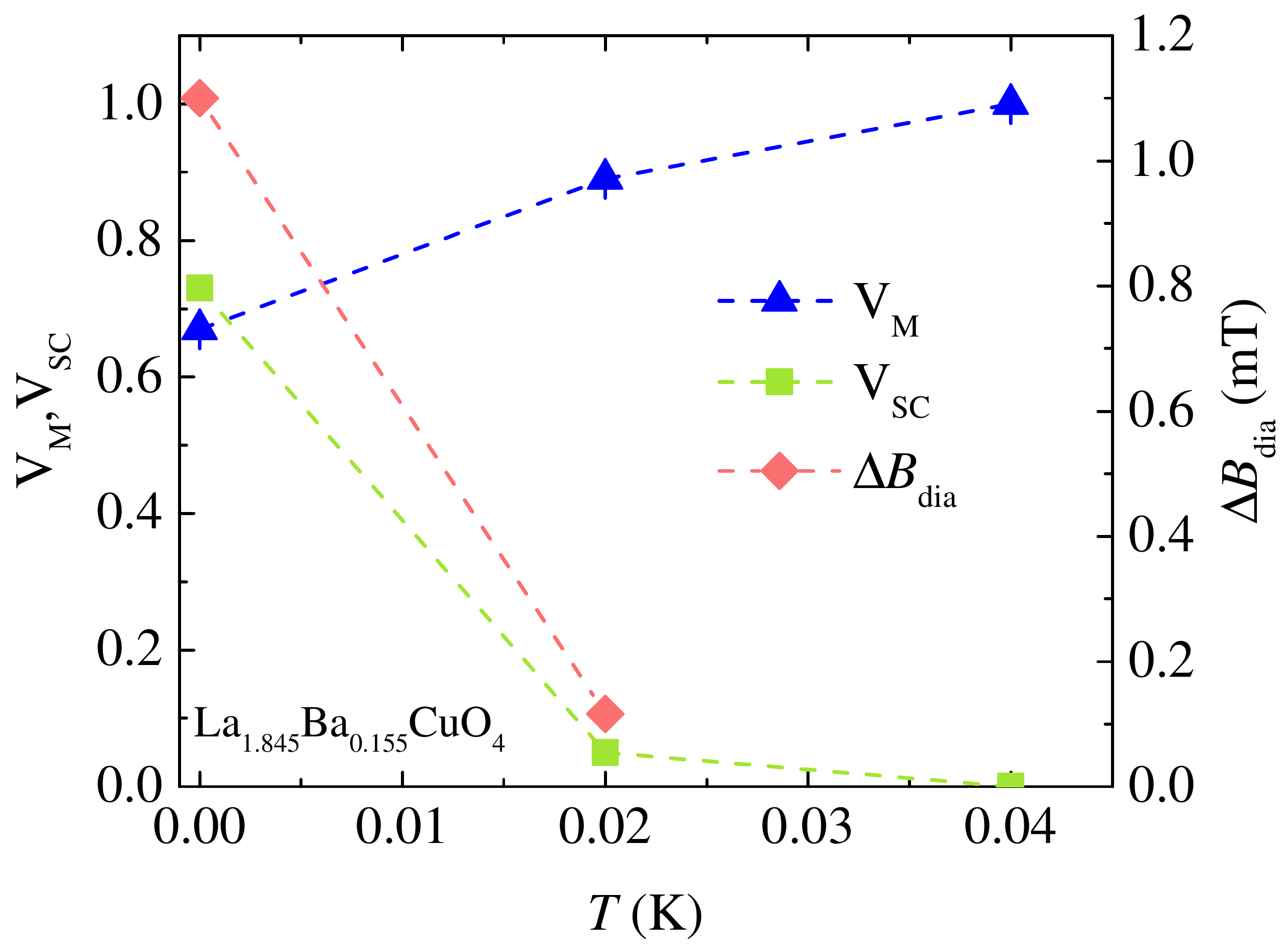}
\vspace{-0.5cm}
\caption{ (Color online) The base temperature values of magnetic volume fraction $V_{\rm m}$, the superconducting volume fraction $V_{\rm SC}$ and the diamagnetic shift ${\Delta}$$B_{{\rm dia}}$ for La$_{1.845}$Ba$_{0.155}$Cu$_{1-y}$Zn$_{y}$O$_{4}$ as a function of Zn-doping. The dashed lines are a guide to the eyes.}  
\label{fig1}
\end{figure}

 Figure 6a shows the temperature dependence of the $c$-axis resistance for the single crystal of La$_{1.48}$Nd$_{0.4}$Sr$_{0.12}$CuO$_{4}$. A clear jump in resistance ($R$) appears at $T_{\rm LTT}$ ${\simeq}$ 71 K, which is indicated by a arrow and ascribed to the structural phase transition from low-temperature orthorhombic (LTO) to low-temperature tetragonal (LTT) phase. Upon lowering the temperature below $T_{\rm LTT}$ ${\simeq}$ 71 K, $R$ monotoneously increases down to 12 K, below which it decreases and reaches zero resistance state towards the base temperature. This indicates that while the onset of SC transition is at $T_{\rm SC}$ in La$_{1.48}$Nd$_{0.4}$Sr$_{0.12}$CuO$_{4}$ is at ${\simeq}$ 12 K, bulk superconductivity is only reached at the base temperature $T_{\rm c}$ ${\simeq}$ 2 K. 

  Figure 6b shows the specific heat $C_{\rm p}$ as a function of temperature for the single crystals of La$_{1.48}$Nd$_{0.4}$Sr$_{0.12}$Cu$_{1-y}$Zn$_{y}$O$_{4}$ ($y$ = 0, 0.016).
Clear peak \cite{Takeda,Crawford} in $C_{\rm p}$ at $T_{\rm LTT}$ ${\simeq}$ 71 K is observed for $y$ = 0 sample, which is in perfect agreement with the resistance data. In Zn-doped sample $y$ = 0.016 the peak appears at slightly lower temperature $T_{\rm LTT}$ ${\simeq}$ 67 K, indicating very tiny impact of the Zn-doping on the bulk LTT structural phase transition temperature.\\

\section{Superconducting properties of La$_{1.845}$Ba$_{0.155}$Cu$_{1-y}$Zn$_{y}$O$_{4}$, studied by means of magnetization and muon-spin rotation experiments}

 Figure 7 shows the temperature dependence of the zero-field-cooled (ZFC) 
diamagnetic moment $m_{\rm ZFC}$ for La$_{1.845}$Ba$_{0.155}$Cu$_{1-y}$Zn$_{y}$O$_{4}$ ($y$ = 0, 0.02, 0.04) sample, recorded in a magnetic field of 
${\mu}_{\rm 0}$$H$ = 0.5 mT. The diamagnetic moment as well as $T_{c}$ are strongly suppressed by Zn-substitution.\

 Figures~8a and b show the temperature dependence of the internal magnetic field $B_{\mu}$ and the magnetic volume fraction $V_{m}$ extracted from the ZF-${\mu}$SR data for La$_{1.845}$Ba$_{0.155}$Cu$_{1-y}$Zn$_{y}$O$_{4}$ ($y$ = 0, 0.02 and 0.04). The low temperature value of $V_{m}$ increases with increasing Zn-content $y$. Note that $V_{m}$ = 100 ${\%}$ for $y$ = 0.04. 
On the other hand, the static spin-stripe order temperature $T_{{\rm so}}$ decreases with increasing $y$ as it is the case for $T_{{\rm c}}$.

 The SC response of the samples was also measured by using TF-${\mu}$SR. 
For all the samples a diamagnetic shift of ${\mu}_{{\rm 0}}$$H_{{\rm int}}$
experienced by the muons is observed below $T_{{\rm c}}$. This
is evident in Fig.~8c where we plot the temperature dependence of
the diamagnetic shift ${\Delta}$$B_{{\rm dia}}$ = ${\mu}_{{\rm 0}}${[}$H_{{\rm int,SC}}$-$H_{{\rm int,NS}}${]}
for La$_{1.845}$Ba$_{0.155}$Cu$_{1-y}$Zn$_{y}$O$_{4}$ ($y$ = 0, 0.02 and 0.04), where ${\mu}_{{\rm 0}}$$H_{{\rm int,SC}}$ denotes the internal field
measured in the SC state and ${\mu}_{{\rm 0}}$$H_{{\rm int,NS}}$ is
the internal field measured in the normal state. Note that ${\mu}_{{\rm 0}}$$H_{{\rm int,NS}}$ is temperature independent.  
The SC transition temperature $T_{{\rm c}}$ is determined from the intercept of the linearly extrapolated ${\Delta}$$B_{{\rm dia}}$
curve and its zero line. The diamagnetic shift ${\Delta}$$B_{{\rm dia}}$ decreases strongly with increasing Zn-content $y$, which is consistent with the reduction of the diamagnetic moment as a result of Zn-substitution. In Fig. 8b of the main text, the values of $T_{{\rm c}}$ and $T_{{\rm so}}$ are plotted as a function of Zn-content $y$. Remarkably, both $T_{{\rm c}}$ and $T_{{\rm so}}$ decrease linearly with increasing $y$, indicating that Zn impurities influence $T_{{\rm c}}$ and $T_{{\rm so}}$  in the same way. These experiments give strong support to our high pressure data, showing the simultaneous occurrence of static magnetism and superconductivity in the LBCO-0.155 system.

Figure 9 shows the Zn-doping evolution of the base temperature values of the magnetic volume fraction, the superconducting volume fraction (estimated from the susceptibility data) and the diamagnetic shift, imposed by the SC state (extracted from ${\mu}$SR). It is clear that while the magnetic fraction is enhanced by Zn-doping, the SC fraction is reduced substantially. Since already for 2 ${\%}$-doped sample the SC fraction is small, it is impossible to extract the reliable information about the superfluid density. Only parameters we can get is the critical temperature and the strength of the diamagnetism.\\

\end{thebibliography}
 

\begin{thebibliography}{22}

\bibitem{Bednorz} Bednorz, J.G., and M\"{u}ller, K.A., $Z.~Phys.~B$ \textbf{64}, 189 (1986).

\bibitem{Moodenbaugh} Moodenbaugh, A.R., Xu, Y., Suenaga, M., Folkerts, T.J., and Shelton, R.N.,
Superconducting properties of La$_{2-x}$Ba$_{x}$CuO$_{4}$. $Phys.~Rev.~B$ \textbf{38}, 4596 (1988).

\bibitem{Kivelson} Kivelson, S.A. $et~al.$
How to detect fluctuating stripes in the high-temperature superconductors.
$Rev.~Mod.~Phys.$ \textbf{75}, 1201 (2003).

\bibitem{Vojta} Vojta, M. 
Lattice symmetry breaking in cuprate superconductors: Stripes, nematics, and superconductivity.
Adv. Phys. \textbf{58}, 699 (2009).

\bibitem{Tranquada1}  Tranquada, J.M., Sternlieb, B.J., Axe, J.D., Nakamura, Y. and Uchida, S., 
Evidence for stripe correlations of spins and holes in copper oxide superconductors.
$Nature~(London)$ \textbf{375}, 561 (1995).

\bibitem{Tranquada2} Tranquada, J.M., Axe, J.D., Ichikawa, N., Nakamura, Y., Uchida, S. and Nachumi, B., 
Neutron-scattering study of stripe-phase order of holes and spins in La$_{1.48}$Nd$_{0.4}$Sr$_{0.12}$CuO$_{4}$.
$Phys.~Rev.~B$ \textbf{54}, 7489 (1996).

\bibitem{Abbamonte} Abbamonte, P. $et~al.$
Spatially modulated 'Mottness' in La$_{2-x}$Ba$_{x}$CuO$_{4}$.
$Nat.~Phys.$ \textbf{1}, 155 (2005).

\bibitem{Huecker} H\"{u}cker, M. $et~al.$
Stripe order in superconducting La$_{2-x}$Ba$_{x}$CuO$_{4}$ (0.095 ${\leq}$ $x$ ${\leq}$ 0.155).
$Phys.~Rev.~B.$ \textbf{83}, 104506 (2011).

\bibitem{Luke} Luke, G.M. $et~al.$ 
Static Magnetic Order in La$_{1.875}$Ba$_{0.125}$CuO$_{4}$.
$Physica~C$ \textbf{185-9}, 1175 (1991).

\bibitem{GuguchiaPRL} Guguchia, Z. $et~al.$ 
Negative Oxygen Isotope Effect on the Static Spin Stripe Order in Superconducting La$_{2-x}$Ba$_{x}$CuO$_{4}$ ($x$ = 1/8) Observed by Muon-Spin Rotation.
$Phys.~Rev.~Lett.$ \textbf{113}, 057002 (2014).


\bibitem{Fujita2004} M. Fujita $et~al.$, ``Stripe order, depinning, and fluctuations in La$_{1.875}$Ba$_{0.125}$CuO$_4$ and La$_{1.875}$Ba$_{0.075}$Sr$_{0.050}$CuO$_4$,'' Phys. Rev. B \textbf{70}, 104517 (2004).

\bibitem{Julien} Wu, T. $et~al.$
Magnetic-field-induced charge-stripe order in the high-temperature superconductor YBa$_{2}$Cu$_{3}$O$_{y}$.
$Nature$ \textbf{477}, 191-194 (2011).

\bibitem{Kohsaka} Kohsaka, Y. $et~al.$ 
An Intrinsic Bond-Centered Electronic Glass with Unidirectional Domains in Underdoped Cuprates.
$Science$ \textbf{315}, 1380 (2007).

\bibitem{Keimer2015}
B. Keimer, S. A. Kivelson, M. R. Norman, S. Uchida, and J. Zaanen, ``From quantum matter to high-temperature superconductivity in copper oxides,'' Nature \textbf{518}, 179 (2015).

\bibitem{Fradkin} Fradkin, E., Kivelson, S. A. and Tranquada, J. M. Colloquium: Theory of intertwined orders in high temperature superconductors. $Rev.~Mod.~Phys.$ \textbf{87}, 457 (2015).

\bibitem{Himeda}  Himeda, A.,  Kato, T. and  Ogata, M. Stripe States with Spatially Oscillating $d$-Wave Superconductivity in the Two-Dimensional t-J Model. $Phys.~Rev.~Lett.$ \textbf{88}, 117001 (2002).

\bibitem{Berg1} Berg, E. $et~al.$
Dynamical Layer Decoupling in a Stripe-Ordered High-$T_{c}$ Superconductor.
$Phys.~Rev.~Lett.$ \textbf{99}, 127003 (2007).


\bibitem{Tranquadareview} Tranquada, J.M. 
Spins, stripes, and superconductivity in hole-doped cuprates.
$AIP~Conference~Proceedings$ \textbf{1550}, 114 (2013).

\bibitem{Tranquada2008} Tranquada, J.M. $et~al.$
Evidence for unusual superconducting correlations coexisting with stripe order in La$_{1.875}$Ba$_{0.125}$CuO$_{4}$.
$Phys.~Rev.~B$ \textbf{78}, 174529 (2008).

\bibitem{Li} Li, Q. $et~al.$ 
Two-Dimensional Superconducting Fluctuations in Stripe-Ordered La$_{1.875}$Ba$_{0.125}$CuO$_{4}$.
$Phys.~Rev.~Lett.$ \textbf{99}, 067001 (2007).

\bibitem{Valla} Valla, T. $et~al.$ 
The Ground State of the Pseudogap in Cuprate Superconductors.
$Science$ \textbf{314}, 1914 (2006).

\bibitem{Shen} He, R.-H. $et~al.$
Energy gaps in the failed high-$T_{c}$ superconductor La$_{1.875}$Ba$_{0.125}$CuO$_{4}$.
$Nat.~Phys.$ \textbf{5}, 119-123 (2009).

\bibitem{Ding2008} J. F. Ding $et~al.$, ``Two-dimensional superconductivity in stripe-ordered 
La$_{1.6-x}$Nd$_{0.4}$Sr$_x$CuO$_4$ single crystals,'' Phys. Rev. B \textbf{77}, 214524 (2008).

\bibitem{Guguchiaarxiv} Guguchia, Z.  $et~al.$,  Cooperative coupling of static magnetism and bulk superconductivity in the
stripe phase of La$_{2-x}$Ba$_{x}$CuO$_{4}$: Pressure ($x$ = 0.155, 0.17) and doping ($x$ = 0.11-0.17) dependent studies.
Phys. Rev. B \textbf{94}, 214511 (2016).

\bibitem{GuguchiaNJP} Guguchia, Z.  $et~al.$
Tuning the static spin-stripe phase and superconductivity in La$_{2-x}$Ba$_{x}$CuO$_{4}$ ($x$ = 1/8) by hydrostatic pressure.
New J. Phys. \textbf{15}, 093005 (2013). 

\bibitem{Xu2014}  Xu, Z. $et~al.$ Neutron-Scattering Evidence for a Periodically Modulated Superconducting Phase in the Underdoped Cuprate La$_{1.905}$Ba$_{0.095}$CuO$_{4}$. $Phys.~Rev.~Lett.$ \textbf{113}, 177002 (2014).

\bibitem{Jakobsen2015}  Jacobsen, H. $et~al.$ 
Neutron scattering study of spin ordering and stripe pinning in superconducting La$_{1.93}$Sr$_{0.07}$CuO$_{4}$.
$Phys.~Rev.~B$ \textbf{92}, 174525 (2015).

\bibitem{Lee} Lee, P. A. Amperean Pairing and the Pseudogap Phase of Cuprate Superconductors.
$Phys.~Rev.~X$ \textbf{4}, 031017 (2014).

\bibitem{Yu} Yu, F. $et~al.$ 
Magnetic phase diagram of underdoped YBa$_{2}$Cu$_{3}$O$_{y}$ inferred from torque magnetization and thermal conductivity.
Proc. Natl. Acad. Sci. \textbf{113}, 12667 (2016).

\bibitem{XiaoG}  Xiao, G., Cieplak, M.Z., Xiao, J.Q., and Chien, C.L., Magnetic pair-breaking effects: Moment formation and critical doping level in superconducting 
La$_{1.85}$Sr$_{0.15}$Cu$_{1-x}$A$_{x}$O$_{4}$ systems ($A$ = Fe, Co, Ni, Zn, Ga, Al).
Phys. Rev. B \textbf{42}, 8752 (1990).

\bibitem{Fukuzumi} Y. Fukuzumi, K. Mizuhashi, K. Takenaka, and S. Uchida.
Universal Superconductor-Insulator Transition and $T_{\rm c}$ Depression in Zn-Substituted
High/$T_{\rm c}$ Cuprates in the Underdoped Regime.
Phys. Rev. Lett. \textbf{76}, 684 (1996). 

\bibitem{Komiya} S. Komiya and Y. Ando,
Electron localization in La$_{2-x}$Sr$_{x}$CuO$_{4}$ and the role of stripes.
Phys. Rev. B \textbf{70}, 060503(R) (2004).

\bibitem{Mendels1994} P. Mendels {\it et al.}, ``Muon-spin-rotation study of the effect of Zn substitution on magnetism in YBa$_2$Cu$_3$O$_x$,'' Phys. Rev. B \textbf{49}, 10035 (1994).

\bibitem{Akoshima2000} Akoshima, M., Koike, Y., Watanabe, I., Nagamine, K., Anomalous muon-spin relaxation in the Zn-substituted YBa$_2$Cu$_{3-2y}$Zn$_{2y}$O$_{6+\delta}$ around the hole concentration of $\frac18$ per Cu.
Phys. Rev. B \textbf{62}, 6761 (2000).

\bibitem{Watanabe2000} Watanabe, I., Akoshima, M., Koike, Y., Ohira, S., Nagamine, K., Muon-spin-relaxation study on the Cu-spin state of Bi$_2$Sr$_2$Ca$_{1-x}$Y$_x$(Cu$_{1-y}$Zn$_y$)$_2$O$_{8+\delta}$ around the hole concentration of $\frac18$ per Cu.
Phys. Rev. B \textbf{62}, 14524 (2000).
 
\bibitem{Adachi2004} Adachi, T., Yairi, S., Koike, Y.,Watanabe, I., Nagamine, K., Muon-spin-relaxation and magnetic-susceptibility studies of the effects of the magnetic impurity 
Ni on the Cu-spin dynamics and superconductivity in La$_{2-x}$Sr$_x$Cu$_{1-y}$Ni$_y$O$_4$ with $x=0.13$. 
Phys. Rev. B \textbf{70}, 060504(R) (2004).

\bibitem{Adachi2004v2} T. Adachi, S. Yairi, K. Takahashi, Y. Koike, I. Watanabe, K. Nagamine, Muon spin relaxation and magnetic susceptibility studies of the effects of nonmagnetic impurities on the Cu spin dynamics and superconductivity in La$_{2-x}$Sr$_x$Cu$_{1-y}$Zn$_y$O$_4$ around $x=0.115$.
Phys. Rev. B \textbf{69}, 184507 (2004).

\bibitem{Koike2012}  Koike, Y., and Adachi, T., Impurity and magnetic field effects on the stripes in cuprates.
Physica C \textbf{481}, 115 (2012).

\bibitem{AnegawaZn} O. Anegawa, Y. Okajima, S. Tanda, and K. Yamaya,
 Effect of spin substitution on stripe order in La$_{1.875}$Ba$_{0.125}$Cu$_{1-y}$M$_{y}$O$_{4}$ ($M$ = Zn or Ni). 
 $Phys.~Rev.~B$ \textbf{63}, 140506 (2001). 

\bibitem{Takeda} J. Takeda, T. Inukai, and M. Sato,
Electronic specific heat of (La,Nd)$_{2-x}$Sr$_{x}$Cu$_{1-y}$Zn$_{y}$O$_{4}$ up to about 300 K.
Journal of Physics and Chemistry of Solids \textbf{62}, 181 (2001).

\bibitem{Fujita} Fujita, M. $et~al.$
Neutron-Scattering Study of Impurity Effect on Stripe Correlations in La-Based 214 High-$T_{\rm c}$ Cuprate.
J. Supercond. Nov. Magn. \textbf{22}, 243 (2009).

\bibitem{suppl}
See Supplemental Material at [URL will be inserted by publisher] for details on the $\mu$SR technique and analysis, and further experimental characterizations of the samples, which includes Refs. \cite{AndreasSuter,Nachumi,Takeda,Crawford}

\bibitem{Eiger} U. Stuhr, B. Roessli, S. Gvasaliya, H.M. R{\o}nnow, U. Filges, D. Graf, A. Bollhalder, D. Hohl, R. B\"urge, M. Schild, L. Holitzner, C. Kaegi, P. Keller and T. M\"uhlebach.
The thermal triple-axis-spectrometer EIGER at the continuous spallation source SINQ.
Nucl. Instrum. Methods Phys.~Res.,~Sect. A \textbf{853}, 16 (2017).

\bibitem{Nachumi} Nachumi, B. $et~al.$
Muon spin relaxation study of the stripe phase order in La$_{1.6-x}$Nd$_{0.4}$Sr$_{x}$CuO$_{4}$ and related 214 cuprates.
$Phys.~Rev.~B$ \textbf{58}, 8760-8772 (1998).

\bibitem{Arai} J. Arai, T. Ishiguro, T. Goko, S. Iigaya, K. Nishiyama, I.
Watanabe, and K. Nagamine, Journal of Low Temperature Physics \textbf{131}, 375 (2003).

\bibitem{AndreasSuter} Suter, A. and Wojek, B.M. 
Musrfit: a free platform-independent framework for ${\mu}$SR data analysis.
$Physics~Procedia$ \textbf{30}, 69-73 (2012).

\bibitem{Crawford} M.K. Crawford, R.L. Harlow, E.M. McCarron, W.E. Farneth, J.D. Axe, H. Chou, and Q.
Huang, Lattice instabilities and the effect of copper-oxygen-sheet distortions on superconductivity in doped La$_{2}$Cu0$_{4}$. Phys. Rev. B \textbf{44}, 749 (1991).

\bibitem{composition}
Note that this is the nominal composition of the crystals, based on the starting materials; the actual composition could differ slightly from that of the polycrystalline samples discussed earlier.

\bibitem{Matsuda2002}
M. Matsuda, M. Fujita, K. Yamada, R. J. Birgeneau, Y. Endoh, and G. Shirane, ``Electronic phase separation in lightly-doped La$_{2-x}$Sr$_x$CuO$_4$,'' Phys. Rev. B \textbf{65}, 134515 (2002).

\bibitem{Sachdev2000}
S. Sachdev, ``Quantum Criticality: Competing Ground States in Low Dimensions,'' Science \textbf{288}, 475 (2000).

\bibitem{Vajk2002}
O. P. Vajk, P. K. Mang, M. Greven, P. M. Gehring, and J. W. Lynn, ``Quantum Impurities in the Two-Dimensional Spin One-Half Heisenberg Antiferromagnet,'' Science \textbf{295}, 1691 (2002).

\bibitem{Mahajan1994}
A. V. Mahajan, H. Alloul, G. Collin, and J. F. Marucco, ``$^{89}$Y NMR Probe of Zn Induced Local Moments in YBa$_2$(Cu$_{1-y}$Zn$_y$)$_3$O$_{6+x}$,'' Phys. Rev. Lett. \textbf{72}, 3100 (1994).

\bibitem{Julien2000}
M.-H. Julien {\it et al.}, ``$^{63}$Cu NMR Evidence for Enhanced Antiferromagnetic Correlations around Zn Impurities in YBa$_2$Cu$_3$O$_{6.7}$,'' Phys. Rev. Lett. \textbf{84}, 3422 (2000).

\bibitem{Birgenau} H. Kimura, K. Hirota, H. Matsushita, K. Yamada, Y. Endoh,
S.-H. Lee, C.F. Majkrzak, R. Erwin, G. Shirane, M. Greven,
Y.S. Lee, M.A. Kastner, and R.J. Birgeneau,
Neutron-scattering study of static antiferromagnetic correlations in La$_{2-x}$Sr$_{x}$Cu$_{1-y}$Zn$_{y}$O$_{4}$. 
Phys. Rev. B \textbf{59}, 6517 (1999).

\bibitem{Balatsky} C.M. Smith, A.H. Castro Neto, and A.V. Balatsky, 
$T_{\rm c}$ suppression in co-doped striped cuprates.
$Phys.~Rev.~Lett.$ \textbf{87}, 177010 (2001). 

\bibitem{Nachumi1996} Nachumi, B. $et~al.$
Muon Spin Relaxation Studies of Zn-Substitution Effects in High-$T_{c}$ Cuprate Superconductors.
$Phys.~Rev.~Lett.$ \textbf{77}, 5421 (1996).

\bibitem{Uemura1989} Y. J. Uemura $et~al.$
Universal Correlations between  $T_{c}$ and $n_{s}$/$m^{*}$  (Carrier Density over Effective Mass) in  High-$T_{c}$ cuprate superconductors.
Phys. Rev. Lett. \textbf{62}, 2317 (1989).
  


\begin{thebibliography}{150}

\bibitem{Nachumi} Nachumi, B. $et~al.$
Muon spin relaxation study of the stripe phase order in La$_{1.6-x}$Nd$_{0.4}$Sr$_{x}$CuO$_{4}$ and related 214 cuprates.
$Phys.~Rev.~B$ \textbf{58}, 8760-8772 (1998).

\bibitem{AndreasSuter} Suter, A. and Wojek, B.M. 
Musrfit: a free platform-independent framework for ${\mu}$SR data analysis.
$Physics~Procedia$ \textbf{30}, 69-73 (2012).

\bibitem{Takeda} J. Takeda, T. Inukai, and M. Sato,
Electronic specific heat of (La,Nd)$_{2-x}$Sr$_{x}$Cu$_{1-y}$Zn$_{y}$O$_{4}$ up to about 300 K.
Journal of Physics and Chemistry of Solids \textbf{62}, 181 (2001).

\bibitem{Crawford} M.K. Crawford, R.L. Harlow, E.M. McCarron, W.E. Farneth, J.D. Axe, H. Chou, and Q. Huang,
Lattice instabilities and the effect of copper-oxygen-sheet distortions on superconductivity in doped La$_{2}$Cu0$_{4}$.
Phys. Rev. B \textbf{44},  749 (1991).



\end{thebibliography}
\end{document}